\documentclass[sigconf]{aamas}

\usepackage{balance} 
\usepackage{subfigure}
\usepackage{enumitem}

\usepackage{graphicx}
\usepackage{caption}
\usepackage{subcaption}
\usepackage{subfigure}
\usepackage{subfloat}
\usepackage{float}

\setcopyright{ifaamas}
\acmConference[AAMAS '24]{Proc.\@ of the 23rd International Conference
on Autonomous Agents and Multiagent Systems (AAMAS 2024)}{May 6 -- 10, 2024}
{Auckland, New Zealand}{N.~Alechina, V.~Dignum, M.~Dastani, J.S.~Sichman (eds.)}
\copyrightyear{2024}
\acmYear{2024}
\acmDOI{}
\acmPrice{}
\acmISBN{}

\acmSubmissionID{949}

\title[AAMAS-2024 Formatting Instructions]{Battlefield Transfers in Coalitional Blotto Games}

\author{Vade Shah}
\affiliation{
  \institution{University of California, Santa Barbara}
  \city{Santa Barbara}
  \country{United States of America}}
\email{vade@ucsb.edu}

\author{Jason R. Marden}
\affiliation{
  \institution{University of California, Santa Barbara}
  \city{Santa Barbara}
  \country{United States of America}}
\email{jrmarden@ece.ucsb.edu}

\begin{abstract}
    In competitive resource allocation environments, agents often choose to form alliances; however, for some agents, doing so may not always be beneficial. Is there a method of forming alliances that always reward each of their members? We study this question using the framework of the coalitional Blotto game, in which two players compete against a common adversary by allocating their budgeted resources across disjoint sets of valued battlefields. On any given battlefield, the agent that allocates a greater amount of resources wins the corresponding battlefield value. Existing work has shown the surprising result that in certain game instances, if one player donates a portion of their budget to the other player, then both players win larger amounts in their separate competitions against the adversary. However, this transfer-based method of alliance formation is not always mutually beneficial, which motivates the search for alternate strategies. In this vein, we study a new method of alliance formation referred to as a joint transfer, whereby players publicly transfer battlefields and budgets between one another before they engage in their separate competitions against the adversary. We show that in almost all game instances, there exists a mutually beneficial joint transfer that strictly increases the payoff of each player.
\end{abstract}


\newcommand{\BibTeX}{\rm B\kern-.05em{\sc i\kern-.025em b}\kern-.08em\TeX}

\newtheorem*{theorem*}{Theorem}

\makeatletter
\gdef\@copyrightpermission{
	\begin{minipage}{0.3\columnwidth}
		\href{https://creativecommons.org/licenses/by/4.0/}{\includegraphics[width=0.90\textwidth]{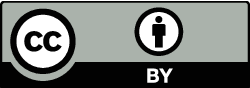}}
	\end{minipage}\hfill
	\begin{minipage}{0.7\columnwidth}
		\href{https://creativecommons.org/licenses/by/4.0/}{This work is licensed under a Creative Commons Attribution International 4.0 License.}
	\end{minipage}
	\vspace{5pt}
}
\makeatother

\begin{document}


\pagestyle{fancy}
\fancyhead{}


\maketitle 


\section{Introduction}

In adversarial resource allocation settings with multiple competitors, agents often seek to form partnerships to obtain some degree of mutual benefit. Sometimes, these alliances are formed through an agreement where agents offer resources to one another. In elections, for example, political parties may exchange financial assets in order to siphon votes from a shared opponent \cite{howard2018ira}. However, in other situations, alliances decide upon some kind of non-compete clause instead. Certain fisheries, for example, establish agreements to decide when and where they can fish, then ally to defend these zones from poachers \cite{de2017deterring, chavez2021endogenous}. Depending on the scenario, one method of alliance formation may be more suitable than another. When considering their odds of success against other opponents, the agents' chosen method of alliance formation is pivotal.

\begin{figure}
    \centering
    \includegraphics[width=\linewidth]{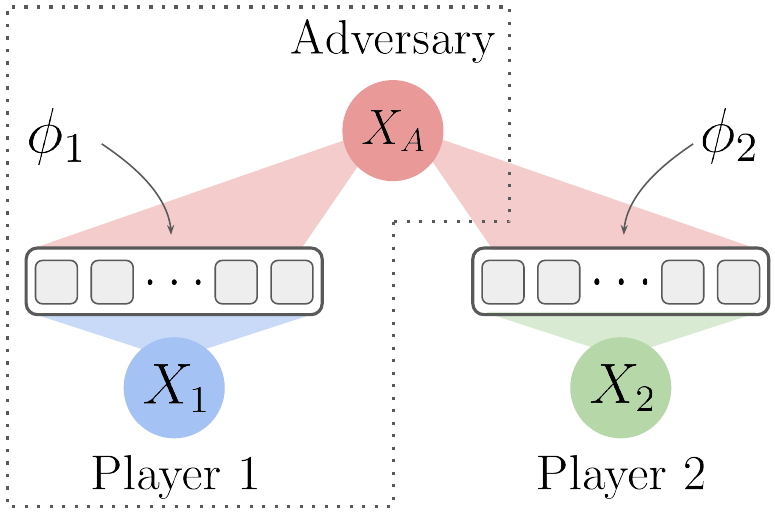}
    \caption{Coalitional Colonel Blotto game depiction. The items enclosed by the dashed box depict a standard Colonel Blotto game between Player 1 and the adversary.}
    \label{fig:coalitional_basic}
\end{figure}

Questions regarding alliances are commonly studied using game theoretic frameworks \cite{bench2009altruism, rahwan2007algorithms}, often with a particular focus on the problem of coalition formation \cite{mahdiraji2021overlapping, jacyno2007understanding} and its applications in fields ranging from defense to computing \cite{yong2003methods, pechoucek2008defence, vorobeychik2015securing}. In this work, we study the potential for collaborative relationships in the context of Colonel Blotto (or General Lotto) games, which offer a well-studied model of strategic resource allocation in adversarial environments. In these games, two competitors allocate their scarce resource budgets across a number of valued battlefields, as shown in the left portion of Figure \ref{fig:coalitional_basic}. On any given battlefield, the agent that has allocated a greater amount of resources wins the associated battlefield value; accordingly, agents distribute their resources across battlefields with the aim of maximizing their accrued value. The framework of the game is general and has been utilized to model strategic behavior in competitive resource allocation settings like political campaigns \cite{behnezhad2018battlefields, thomas2018n} and cybersecurity threats \cite{ferdowsi2017colonel, gupta2014three}.

To analyze methods of partnership formation, we utilize the framework of the \textit{coalitional} Colonel Blotto game, through which we seek to identify when opportunities emerge for different types of collaborations between players. A foundational model of the coalitional Blotto game is given in \cite{kovenock2012coalitional}, where two players compete in disjoint Blotto games against a common adversary as depicted in Figure \ref{fig:coalitional_basic}. Their work demonstrates a seemingly counterintuitive result: In some scenarios, if one of the two players (say, Player 1) donates a portion of their budget to the other player (Player 2), both players experience greater payoffs in their individual games against the adversary. That is, Player 1 becomes weaker, Player 2 becomes stronger, but in their separate competitions against the adversary, each of them win more battlefield value than they would have if no such budgetary transfer had occurred. The main contribution of \cite{kovenock2012coalitional} is a complete characterization of all coalitional Colonel Blotto games in which there exist budgetary transfers that strictly benefit Players 1 and 2.

However, their characterization shows that opportunities for mutual improvement with budgetary transfers are limited; in a nontrivial subset of game instances, any possible budgetary transfer strictly decreases one player's payoff. As such, many studies have explored other strategic opportunities for players to gain a competitive advantage in the same two-versus-one setting, each with different variations on the actions available to the players and the information structure of the game. In the existing literature, scholars have analyzed alternate forms of budgetary transfers \cite{kovenock2012coalitional, heyman2018colonel}, battlefield additions \cite{kovenock2010asymmetric}, informational asymmetries \cite{chandan2020showing, paarporn2019characterizing}, and various combinations of these strategies and structures \cite{gupta2014three, gupta2014provide}. In this work, we consider a new strategic maneuver: \textit{joint transfers}. In a joint transfer, players can transfer their budgets \textit{and} battlefields simultaneously. When performing a so-called \textit{battlefield transfer}, one player abstains from competing on a subset of their battlefields, allowing the other player to compete against the adversary for these battlefields instead. Battlefield transfers, to the best of our knowledge, have not received attention in the literature.

To better understand the concept of battlefield transfers, let us study an example. Consider an election in which Party 1 and Party 2 expend campaign funds across disjoint sets of voting districts in competition against a common adversarial party; each party campaigns with the goal of winning as many seats as possible across all voting districts. One might imagine a scheme whereby Party 1 abstains from campaigning for a subset of seats and allows Party 2 to campaign for them instead, thereby \textit{transferring districts} to Party 2. However, whether such a transfer strategy could ever be mutually beneficial is not immediately evident. Moreover, even if it could be beneficial, it is not clear exactly when such an opportunity might arise or how it might compare to other strategic maneuvers.

\begin{figure*}
    \includegraphics[width=\textwidth]{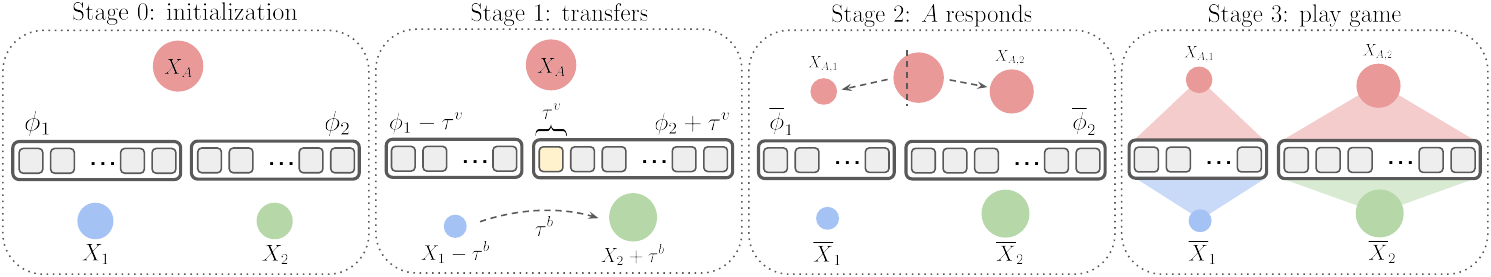}
    \caption{The stages of the coalitional Blotto game. In Stage 0, the game is initialized. In Stage 1, the two players publicly perform mutually beneficial transfers. In Stage 2, the adversary determines how to optimally allocate their budget to each competition. In Stage 3, the two disjoint Blotto games are played.}
    \label{fig:coalitional_full}
\end{figure*}

To answer these questions regarding the existence and nature of mutually beneficial joint transfers, we adopt the coalitional Blotto framework presented in \cite{kovenock2012coalitional}. The model, as shown in Figure \ref{fig:coalitional_full}, describes a three-stage coalitional Colonel Blotto game. The setup of the game is as follows. Two players, who are not necessarily allies, compete against a common adversary. Before the game begins, it is initialized in a so-called Stage 0. The players, referred to as Players 1 and 2, are equipped with budgets $X_1$ and $X_2$, respectively, and the adversary is equipped with a budget $X_A$. Player $i \in \{ 1, 2 \}$ competes against the adversary for battlefields that have a total valuation of $\phi_i$, and before the game begins, the values of the battlefields and the agents' budgets are made publicly known. The game starts in Stage 1, when Player 1 transfers a portion of their budget $\tau^b$ and a portion of their battlefield valuation $\tau^v$ to Player 2. They perform this joint transfer only if it would be mutually beneficial, meaning that each player would acquire greater value in their separate competitions against the adversary than they would have if no transfer occurred. In Stage 2, having observed any transfers, the common adversary decides how to best allocate their resources towards each of their games against Players 1 and 2, i.e., they choose $X_{A,1}$, $X_{A,2}$ such that their payoff is maximized, where $X_{A,1} + X_{A,2} \leq X_A$. Lastly, in Stage 3, all of the agents deploy their resources, the disjoint Blotto games are played, and the payoffs are realized.

Within this setting, our work seeks to address the following question:

\begin{quote}
    Are there increased opportunities for mutual improvement if players can transfer budgets \textit{and} battlefields?
\end{quote}

We answer this question in the affirmative in our main Theorem. In particular, we show that for almost all game instances, there exists a simultaneous transfer of battlefields and budgets that is beneficial for each player, and we fully characterize the subset of game instances where such an opportunity is available. The results lend new insight into the nature of alliance formation through the framework of the coalitional Blotto game.


\section{Model}

For ease of introduction, we begin our discussion with a brief yet formal introduction of the standard Colonel Blotto game. Then, we present the model of the coalitional Colonel Blotto game adopted in this work.

\subsection{Standard Colonel Blotto Game}

A standard Colonel Blotto game is shown in the left half of Figure \ref{fig:coalitional_basic}. In the game, two agents (Player 1 and the adversary) expend their resources ($X_1$ and $X_A$) across a set of battlefields with the aim of maximizing their accrued valuation. On each battlefield, the agent that allocates a greater level of resources will win that battlefield's valuation\footnote{In the event of the tie, we assume that the the agents share the payoff equally. However, when agents follow their equilibrium strategies, a tie is a measure zero event, so the choice of tie-breaking rule does not affect the equilibrium payoffs.}. To simplify the analysis, we assume that the number of battlefields is arbitrarily large, which allows us to analyze the game with regards to the \textit{total} valuation across all battlefields (here, $\phi_1$) rather than the value of each individual battlefield.

When determining how to allocate their resources, we assume that each agent has full knowledge of the battlefield valuations as well as their opponent's budget. The standard game is a one-shot game; Player 1 and the adversary simultaneously allocate their budgets to the battlefields, and their payoffs are realized accordingly.

The agents' equilibrium strategies\footnote{Throughout the text, we assume that each agent allocates their budgets to battlefields according to their equilibrium strategies. For a complete characterization and discussion of their equilibrium strategies, we refer the reader to \cite{kovenock2012coalitional}.} and payoffs in the standard Colonel Blotto game are characterized in \cite{roberson2006colonel}, which we summarize below. If $X_1 \geq X_A$, then the equilibrium payoffs of the adversary and Player 1 are given by
\begin{align*}
    u_A(\phi_1, X_1, X_A) &= \phi_1 \left( \frac{X_A}{2 X_1} \right), \\
    \quad u_1(\phi_1, X_1, X_A) &= \phi_1 \left( 1 - \frac{X_A}{2 X_1} \right),
\end{align*}
respectively. When $X_1 < X_A$, the payoffs are obtained by swapping the subscripts $1$ and $A$ in the equations above.

Equipped with this understanding of the standard game, we now turn our attention to the coalitional game.

\subsection{Coalitional Colonel Blotto Game}

The three-agent, three-stage, full-information coalitional Colonel Blotto game is depicted in Figure \ref{fig:coalitional_basic}. The game proceeds in the following stages.

\subsubsection*{Stage 0} The game is initialized. Two players, henceforth referred to as Players 1 and 2, compete with a common adversary on disjoint sets of battlefields. In essence, each player competes in a standard Blotto game against the adversary; Player $i$ uses their budget $X_i$ to compete across a set of battlefields with total valuation $\phi_i$. We assume that the budgets are normalized so that $X_A = 1$. Thus, a coalitional Blotto game instance $G$ is fully parameterized by $G = (\phi_1, \phi_2, X_1, X_2) \in \mathcal{G} = \mathbb{R}_{>0}^4$.

\subsubsection*{Stage 1} In the first stage, Players 1 and 2 decide whether to engage in a joint transfer. Formally, we define a joint transfer as a tuple $\tau \triangleq (\tau^b, \tau^v) \in (-X_1, X_2) \times (-\phi_2, \phi_1)$, where $\tau^b \in (-X_1, X_2)$ is the net amount of budget transferred from Player 1 to 2 (a negative $\tau^b$ indicates that the net transfer goes in the opposite direction), and $\tau^v \in (-\phi_2, \phi_1)$ is the net amount of  battlefield valuation transferred from Player 1 to 2 (a negative $\tau^v$ indicates that the net transfer goes in the opposite direction). Thus, the post-transfer budgets of Players 1 and 2 are given by $\overline{X}_1 \triangleq X_1 - \tau^b$ and $\overline{X}_2 \triangleq X_2 + \tau^b$, respectively, and the post-transfer battlefield valuations of Players 1 and 2 are given by $\overline{\phi}_1 \triangleq \phi_1 - \tau^v$ and $\overline{\phi}_2 \triangleq \phi_2 + \tau^v$, respectively.

Using the definition of joint transfers, it is straightforward to define budgetary transfers as particular instances of joint transfers in which $\tau^b \neq 0$ and $\tau^v = 0$. Similarly, battlefield transfers are defined as particular instances of joint transfers in which $\tau^v \neq 0$ and $\tau^b = 0$. To simplify the subsequent analysis, we assume that battlefields are arbitrarily divisible so that any amount of battlefield valuation in $(-\phi_2, \phi_1)$ can be feasibly transferred (i.e., entire battlefields need not be transferred). Throughout the text, the terms 'battlefield transfer' and 'valuation transfer' will be used interchangeably.

A joint transfer is depicted in the middle left panel of Figure \ref{fig:coalitional_full}. Player 1's budget decreases while Player 2's budget increases as the result of a budgetary transfer. Similarly, the total valuation across Player 1's battlefields decreases while the total valuation across Player 2's battlefields increases as the result of a battlefield transfer (depicted as a yellow square).

We assume that in Stage 1, players only perform transfers that are mutually beneficial, i.e., the transfer satisfies \eqref{eq:ben}. Any transfer that occurs in this stage is binding, and the post-transfer battlefield valuations and budgets become publicly known. Here, we find it important to note that since the players’ decision to perform a transfer is determined jointly, not individually, the notion of a 'Nash equilibrium' transfer is not well-defined.

\subsubsection*{Stage 2}

In the second stage, after having observed all transfers between the players, the common adversary decides how to allocate their budget to their games against Players 1 and 2. This is depicted in the middle right panel of Figure \ref{fig:coalitional_full}. We refer to the adversary's allocation to their game against Player $i$ as $X_{A, i} (G, \tau)$ (written as $X_{A,i}$ for brevity when the dependence is clear), where $X_{A, 1} + X_{A, 2} \leq X_A$. The adversary chooses their allocation such that their payoff $u_A(G, \tau)$ is maximized. Their optimal allocation strategy for a given game and transfer is derived in \cite{kovenock2012coalitional}, which we summarize in Table \ref{table1}.

\subsubsection*{Stage 3}

In the third and final stage, all three agents allocate their resources to their respective battlefields, the two separate standard Blotto games are played, and the agents' payoffs are realized. This is depicted in the rightmost panel of Figure \ref{fig:coalitional_full}. The equilibrium payoffs of the players in the coalitional Blotto game \cite{kovenock2012coalitional} are given by
\begin{equation}\label{eq:payoff}
    \begin{split}
        u_1(G, \tau) &= \begin{cases}
            (\phi_1 - \tau^v) \left( \frac{X_1 - \tau^b}{2 X_{A, 1}(G, \tau)} \right) & X_1 \leq X_{A, 1} \\
            (\phi_1 - \tau^v) \left( 1 - \frac{X_{A, 1}(G, \tau)}{2 (X_1 - \tau^b)} \right) & \text{otherwise},
        \end{cases} \\
        u_2(G, \tau) &= \begin{cases}
            (\phi_2 + \tau^v) \left( \frac{X_2 + \tau^b}{2 X_{A, 2}(G, \tau)} \right) & X_2 \leq X_{A, 2} \\
            (\phi_2 + \tau^v) \left( 1 - \frac{X_{A, 2}(G, \tau)}{2 (X_2 + \tau^b)} \right) & \text{otherwise},
        \end{cases}
    \end{split}
\end{equation}
and the equilibrium payoff of the adversary is given by
\begin{equation*}
    u_A(G, \tau) = \phi_1 + \phi_2 - u_1(G, \tau) - u_2(G, \tau).
\end{equation*}

The focus of this paper is on understanding the subset of game instances (i.e., the subset of the parameter space $\mathcal{G}$) in which there exists any mutually beneficial joint transfer. Specifically, we call a transfer $\tau$ \textit{mutually beneficial} if and only if
\begin{equation}\label{eq:ben}
    u_i(G, \tau) > u_i(G, 0) \quad \forall \; i \in \{1, 2\},
\end{equation}
where, with a slight abuse of notation, $u_i(G, 0)$ denotes Player $i$'s payoff when no transfers occur.

\subsection{Numerical Example}\label{section:example}

Before proceeding to the main results of the paper, we briefly discuss two numerical examples that lend insight into the nature of mutually beneficial transfers.

\begin{figure}
    \centering
    \includegraphics[width=\linewidth]{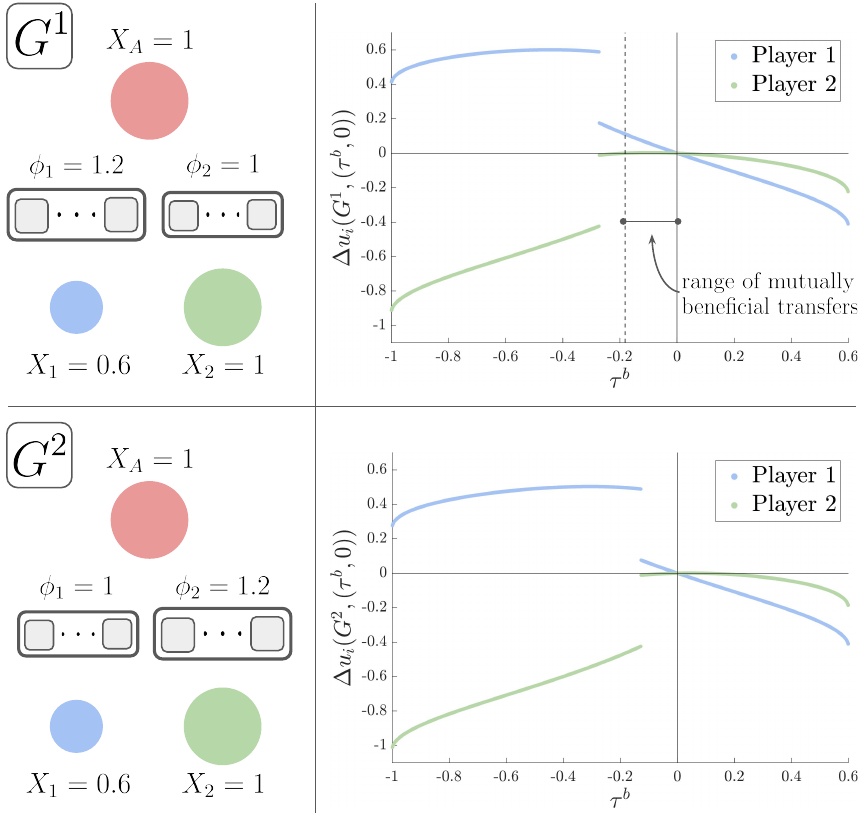}
    \caption{Cartoon depictions of games $G^1$ and $G^2$, along with the change in payoff of each player as a function of the budgetary transfer for each game.}
    \label{fig:example}
\end{figure}

First, consider the game $G^1 = (1.2, 1, 0.6, 1)$, graphically depicted in the top-left portion of Figure \ref{fig:example}. For this game, the result from \cite{kovenock2012coalitional} asserts that there must exist a mutually beneficial budgetary transfer. The plot in the top-right of Figure \ref{fig:example} shows the change in payoff of Player $i$, $\Delta u_i(G, \tau) \triangleq u_i(G, \tau) - u_i(G, 0)$, as a function of the budgetary transfer $\tau^b$. There is indeed a range of budgetary transfers that strictly increase the payoff of each player.

However, consider a similar game $G^2 = (1, 1.2, 0.6, 1)$ depicted in the bottom-left of Figure \ref{fig:example}. Notice that the only difference between $G^1$ and $G^2$ is that the battlefield valuations in the two games are swapped. Despite the similarity between the two game instances, there is no opportunity for mutually beneficial budgetary transfers in $G^2$. This is shown in the bottom-right plot in Figure \ref{fig:example}, which shows the relative change in payoff for each player as a function of the budgetary transfer amount $\tau^b$. Notice that for every possible transfer value $\tau^b \neq 0$,  at least one player receives a strictly lower payoff.

\section{Results}

In this section, we present our main contribution. We characterize $\mathcal{G}_0^{b,v}$, the subset of $\mathcal{G}$ in which mutually beneficial joint transfers do not exist, and we show that this set is of measure zero. In doing so, we show that $\mathcal{G}^{b,v}$, the subset of $\mathcal{G}$ in which mutually beneficial joint transfers do exist, is of full measure, meaning that there are almost always opportunities for mutual improvement when players consider joint transfers. After proving these statements, we briefly explore their implications in another numerical example.

We begin with the statement of our main Theorem.

\begin{theorem*}\label{thm:joint}
    Let $\mathcal{G}^{b,v}_0 \in \mathcal{G}$ denote the subset of all coalitional Blotto games for which there do not exist any mutually beneficial joint transfers. Then, $\mathcal{G}^{b,v}_0$ has measure zero.
\end{theorem*}

\begin{figure*}
    \includegraphics[width=\textwidth]{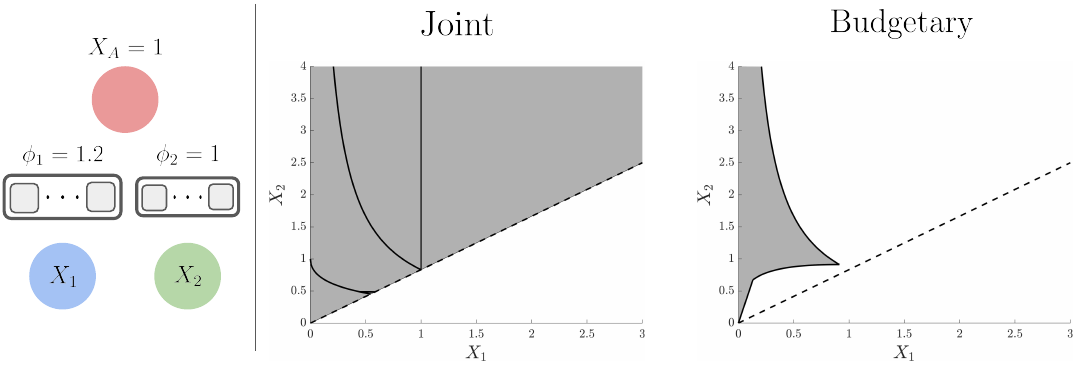}
    \caption{For a game with initial battlefield valuations $\phi_1 = 1.2$ and $\phi_2 = 1$ (depicted on the left), plots of the subsets of the parameter space in which mutually beneficial joint transfers (middle) and budgetary transfers (right) exist. The budgets of Players 1 and 2 represent the horizontal and vertical axes, respectively, of each of the plots. The shaded regions indicate where there exists a mutually beneficial transfer. Note that the subsets are depicted only for parameters satisfying $\frac{\phi_1}{\phi_2} \geq \frac{X_1}{X_2}$ to avoid redundancy.}
    \label{fig:regions}
\end{figure*}

The subset of games for which joint transfers are mutually beneficial is highlighted in Figure \ref{fig:regions}. As shown in the left plot, a mutually beneficial joint transfer exists almost everywhere in the parameter space $\mathcal{G}$; in particular, they exist everywhere except along the solid black curves, which depict the measure-zero subset $G_0^{b, v}$. Notice the contrast between $\mathcal{G}^{b, v}$ and $\mathcal{G}^b$, the subset of the parameter space in which there exist mutually beneficial budgetary transfers, shown in the right plot in Figure \ref{fig:regions}. There are large subsets of the parameter space for which there are no opportunities for mutual improvement when players are limited to budgetary transfers, but if they are allowed to perform budgetary and battlefield transfers simultaneously, then there is an opportunity for mutual improvement in almost all game instances.

Let us now proceed to the proof of the Theorem.

\subsection{Proof of Theorem}\label{section:thm_proof}

The optimal allocation of the adversary takes one of four qualitatively distinct forms depending on the parameters of the game, which are summarized in Table \ref{table1}. The proof proceeds by analyzing the adversary's optimal allocation and the resulting payoffs of the players for each of the four forms. Note that the following analysis is conducted under the assumption that
\begin{equation}
    \frac{\overline{\phi}_1}{\overline{\phi}_2} > \frac{\overline{X}_1}{\overline{X}_2}, \label{ineq:all}
\end{equation}
but the arguments are identical when the opposite is true, as they only require a swapping of indices.

\begin{table}[h]
\centering
\caption{The adversary's optimal budget allocation}
\label{table1}
\begin{center}
\begin{tabular}{|c|c|c|}
\hline
    Case & Condition & $X_{A, 1}$ \\
    \hline
    1 & $ \frac{\overline{\phi}_1}{\overline{\phi}_2} > \frac{\overline{X}_1}{\overline{X}_2} \text{ and } \frac{\overline{\phi}_1}{\overline{\phi}_2} \geq \frac{1}{\overline{X}_1 \overline{X}_2} $ & $1$ \\
    \hline
    2 & $\frac{\overline{\phi}_1}{\overline{\phi}_2} > \frac{\overline{X}_1}{\overline{X}_2}$ and $0 < 1 - \sqrt{\frac{\overline{\phi}_i \overline{X}_1 \overline{X}_2}{\overline{\phi}_{-i}}} \leq \overline{X}_2$ & $\sqrt{\frac{\overline{\phi}_i \overline{X}_1 \overline{X}_2}{\overline{\phi}_{-i}}}$ \\
    \hline
    3 & $\frac{\overline{\phi}_i}{\overline{\phi}_{-i}} \geq \frac{\overline{X}_1}{\overline{X}_2}$ and $1 - \sqrt{\frac{\overline{\phi}_i \overline{X}_1 \overline{X}_2}{\overline{\phi}_{-i}}} > \overline{X}_2$ & $\frac{\sqrt{\overline{\phi}_i \overline{X}_1}}{\sqrt{\overline{\phi}_i \overline{X}_1} + \sqrt{\overline{\phi}_{-i} \overline{X}_2}}$ \\
    \hline
    4 & $\frac{\overline{\phi}_i}{\overline{\phi}_{-i}} = \frac{\overline{X}_1}{\overline{X}_2}$ and $1 \leq \overline{X}_1 + \overline{X}_2$ & $X_{A, i} \leq X_i$ \\
    \hline
\end{tabular}
\end{center}
\end{table}

Before proceeding, we briefly describe each of the four cases in Table \ref{table1} below.

\begin{itemize}
    \item \textbf{Case 1:} The adversary allocates all of their resources against Player 1, who is poorly endowed in comparison to Player 2.
    \item \textbf{Case 2:} The adversary reaches a point of diminishing returns in their game against Player 1 and thus diverts some of their resources towards their game against Player 2. 
    \item \textbf{Case 3:} The adversary has a larger budget than Players 1 and 2 combined (i.e., $X_1 + X_2 < 1$), so the adversary allocates their budget such that the marginal payoffs in each game are equal. 
    \item \textbf{Case 4:} The adversary's marginal payoff in each game is equal so long as their allocation to each game is less than the corresponding player's budget.
\end{itemize}

We will now analyze the players' payoff functions in each of these four comprehensive cases.

\renewcommand{\arraystretch}{1}
\subsubsection*{Case 1}
The first case addresses any game whose pre- and post-transfer parameters strictly satisfy the conditions of Case 1 provided in Table \ref{table1}. In this case, the optimal response of the adversary is to allocate the entirety of their budget towards their game against Player 1, i.e., $X_{A, 1} = 1$, $X_{A, 2} = 0$. Thus, according to (\ref{eq:payoff}), the payoff functions of each player are given by
\begin{align*}
    u_1(G, \tau) &= 
    \begin{cases}
        (\phi_1 - \tau^v) \left( \frac{X_1 - \tau^b}{2} \right) & X_1 - \tau^b \leq 1 \\
        (\phi_1 - \tau^v) \left( 1 - \frac{1}{2 (X_1 - \tau^b)} \right) & \text{otherwise}, 
    \end{cases} \\
    u_2(G, \tau) &= \phi_2 + \tau^v.
\end{align*}
First, observe that the payoff functions are locally continuous and differentiable when $X_1 - \tau^b \neq 1$, so their gradients are well-defined. Evaluated at $\tau = 0$, the gradients of each function with respect to $\tau$ are given by
\begin{gather*}
    \nabla u_1(G, 0) = 
    \begin{cases}
        \begin{bmatrix} -\frac{\phi_1}{2} \\ -\frac{X_1}{2} \end{bmatrix} & X_1 < 1 \\
        \begin{bmatrix} -\frac{\phi_1}{2 X_1^2} \\ \frac{1}{2 X_1} - 1 \end{bmatrix} & X_1 > 1,
    \end{cases} \quad \text{ and } \quad \nabla u_2(G, 0) = \begin{bmatrix} 0 \\ 1 \end{bmatrix}.
\end{gather*}
Recall that if the gradients of two functions $u_1$ and $u_2$ are not diametrically opposed, i.e., if
\begin{equation*}
    \nabla u_1(G, 0) \neq -\beta \nabla u_2(G, 0)
\end{equation*}
for any $\beta \in \mathbb{R}_{>0}$, then there exists some direction $\tau^*$ such that, by the definition of the gradient, $u_1$ and $u_2$ increase along $\tau^*$. Thus, all that remains to be shown is that the gradients of the payoff functions are not diametrically opposed. We proceed by contradiction. Suppose that $\nabla u_1(G, 0) = -\beta \nabla u_2(G, 0)$. This would imply $\phi_1 = 0$, which contradicts the assumption that the initial battlefield valuations are strictly positive. Thus, we define $\mathcal{Z}_0^1$ as
\begin{align*}
    \mathcal{Z}_0^1 \triangleq &\{G \in \mathcal{G} \vert X_1 \neq 1 \text{ and } G \text{ belongs to Case 1} \},
\end{align*}
which is a measure-zero subset of $\mathcal{G}$.

\subsubsection*{Case 2}
Next, consider any game whose pre- and post-transfer parameters strictly satisfy the conditions of Case 2 provided in Table \ref{table1}. In this case, the optimal response of the adversary is to split their budget such that $X_{A, 1} = \left( \frac{\overline{\phi_1} \overline{X_1} \overline{X_2}}{\overline{\phi_2}} \right)^{\frac{1}{2}}$ and $X_{A, 2} = 1 - X_{A, 1}$. Thus, the payoff functions of each player are given by
\begin{align*}
    u_1(G, \tau) &= \frac{1}{2} \left( \frac{(X_1 - \tau^b)(\phi_1 - \tau^v) (\phi_2 + \tau^v)}{X_2 + \tau^b} \right)^{\frac{1}{2}}, \\
    u_2(G, \tau) &= (\phi_2 + \tau^v) \left( 1 - \frac{1}{2 (X_2 + \tau^b)} \right) \\
    &+ \frac{1}{2} \left( \frac{(X_1 - \tau^b)(\phi_1 - \tau^v) (\phi_2 + \tau^v)}{X_2 + \tau^b} \right)^{\frac{1}{2}}
\end{align*}
Evaluated at $\tau = 0$, the gradients of each function are given by
\begin{align*}
    \nabla u_1(G, 0) &= \begin{bmatrix}
    -\frac{(\phi_1 \phi_2)^{\frac{1}{2}} (X_1 + X_2)}{4 X_2^{\frac{3}{2}} X_1^{\frac{1}{2}}} \\
    \frac{1}{4} (\phi_1 - \phi_2) \left( \frac{X_1}{X_2 \phi_1 \phi_2} \right)^{\frac{1}{2}}
    \end{bmatrix}, \\
    \nabla u_2(G, 0) &= \begin{bmatrix}
    \frac{\phi_2}{2 X_2^2} - \frac{(\phi_1\phi_2)^{\frac{1}{2}} (X_1 + X_2)}{4 X_2^{\frac{3}{2}} X_1^{\frac{1}{2}}} \\
    1 - \frac{1}{2 X_2} + \frac{1}{4} (\phi_1 - \phi_2) \left( \frac{X_1}{X_2 \phi_1 \phi_2} \right)^{\frac{1}{2}}
    \end{bmatrix}.
\end{align*}
Observe that $\nabla u_1$ is always nonzero; however, $\nabla u_2$ is equal to 0 if the parameters of $G$ are such that 
\begin{equation}\label{eq:case_2_max}
    (X_1 + X_2)(\phi_1 - \phi_2) = \frac{4 \phi_2 (1 - 2 X_2)}{X_2}.
\end{equation}
This implies that there may not exist any opportunity for an arbitrarily small mutually beneficial joint transfer for a game whose parameters satisfy \eqref{eq:case_2_max}, since the utility of Player 2 is at an extremum. Recall that assessing the type of the extremum requires characterizing the Hessian of $u_2$, which is a tedious task; however, we disregard this analysis, as the subset of games that satisfy \eqref{eq:case_2_max} is in fact a measure-zero subset of $\mathcal{G}$. Next, we analyze the condition of diametric opposition, i.e., $\nabla u_1(G, 0) \neq -\beta \nabla u_2(G, 0).$  To simplify the subsequent algebraic manipulation, we first consider the case where $\phi_1 = \phi_2$, which implies $\frac{\partial u_1}{\partial \tau_v} \big\vert_0 = 0$. In this case, the gradients may be diametrically opposed if  $\frac{\partial u_2}{\partial \tau_v} \big\vert_0 = 0$ also, which occurs when $X_2 = \frac{1}{2}$. However, if $\phi_1 = \phi_2$, then $X_2 > \frac{1}{2}$ for any game belonging to Case 2, so this case is not of concern. Hence, assuming $\phi_1 \neq \phi_2$, it is straightforward to show that the condition of diametric opposition can be rewritten as
\begin{equation}\label{eq:case_2_opp}
    \frac{\phi_1}{\phi_1 - \phi_2} (1 - 2 X_2) = \frac{X_1}{X_1 + X_2}.
\end{equation}
Thus, we define $\mathcal{Z}_0^2$ as
\begin{align*}
    \mathcal{Z}_0^2 \triangleq &\{G \in \mathcal{G} \vert G \text{ belongs to Case 2} \} \\ 
    \cap &\{G \in \mathcal{G} \vert G \text{ satisfies \eqref{eq:case_2_max} or \eqref{eq:case_2_opp}} \}
\end{align*}
which is a measure-zero subset of $\mathcal{G}$.

\subsubsection*{Case 3}
Next, consider any game whose pre- and post-transfer parameters strictly satisfy the conditions of Case 3 provided in Table \ref{table1}. In this case, the optimal response of the adversary is to split their budget such that $X_{A, 1} = \frac{(\overline{\phi}_1 \overline{X}_1)^{\frac{1}{2}}}{(\overline{\phi}_1 \overline{X}_1)^{\frac{1}{2}} + (\overline{\phi}_2 \overline{X}_2)^{\frac{1}{2}}}$ and $X_{A, 2} = 1 - X_{A, 1}$. Thus, the payoff functions of each player are given by
\begin{align*}
    u_1(G, \tau) = &\frac{1}{2} (\phi_1 - \tau^v) (X_1 - \tau^b) \\
    + &\frac{1}{2} ((X_1 - \tau^b)(X_2 + \tau^b)(\phi_2 + \tau^v)(\phi_1 - \tau^v))^{\frac{1}{2}}, \\
    u_2(G, \tau) = &\frac{1}{2} (\phi_2 + \tau^v)(X_2 + \tau^b) \\
    + &\frac{1}{2} ((X_1 - \tau^b)(X_2 + \tau^b)(\phi_1 - \tau^v)(\phi_2 + \tau^v))^{\frac{1}{2}}
\end{align*}
Evaluated at $\tau = 0$, the gradients of each function are given by
\begin{align*}
    \nabla u_1(G, 0) &= \begin{bmatrix} 
    -\frac{1}{2}\phi_1 + \frac{1}{4} \left( \frac{\phi_1 \phi_2}{X_1 X_2} \right)^{\frac{1}{2}} (X_1 - X_2) \\
    -\frac{1}{2} X_1 + \frac{1}{4} \left( \frac{X_1 X_2}{\phi_1 \phi_2} \right)^{\frac{1}{2}} (\phi_1 - \phi_2)
    \end{bmatrix}, \\
    \nabla u_2(G, 0) &= \begin{bmatrix} 
    \frac{1}{2}\phi_2 + \frac{1}{4} \left( \frac{\phi_1 \phi_2}{X_1 X_2} \right)^{\frac{1}{2}} (X_1 - X_2) \\
    \frac{1}{2} X_2 + \frac{1}{4} \left( \frac{X_1 X_2}{\phi_1 \phi_2} \right)^{\frac{1}{2}} (\phi_1 - \phi_2)
    \end{bmatrix}.
\end{align*}
Observe that each function can attain an extremum; specifically, when
\begin{align}
    (X_1 - X_2) (\phi_1 - \phi_2) &= 4 X_1 \phi_1, \label{eq:case_3_max_1}
    \intertext{$\nabla u_1(G, 0) = 0$, and when}
    (X_1 - X_2) (\phi_1 - \phi_2) &= 4 X_2 \phi_2, \label{eq:case_3_max_2}
\end{align}
$\nabla u_2(G, 0) = 0$. Once again, characterizing the Hessians of each function is unnecessary because the subset of games that satisfy \eqref{eq:case_3_max_1} or \eqref{eq:case_3_max_2} is a measure-zero subset of $\mathcal{G}$. We can now analyze the condition of diametric opposition, i.e., $\nabla u_1(G, 0) \neq -\beta \nabla u_2(G, 0)$. Observe that $\frac{\partial u_1}{\partial \tau_b} \big\vert_0 = 0 \implies \frac{\partial u_2}{\partial \tau_b} \big\vert_0 \neq 0$, and similarly, $\frac{\partial u_1}{\partial \tau_v} \big\vert_0 = 0 \implies \frac{\partial u_2}{\partial \tau_v} \big\vert_0 \neq 0$, and the same holds true when the subscripts are swapped; thus, the gradients cannot be diametrically opposed when any partial derivative is $0$. Having addressed this case, we can assume all of the partial derivatives are nonzero. It is straightforward to show that the condition of diametric opposition can be rewritten as
\begin{equation}\label{eq:case_3_opp}
    X_2 = \frac{\phi_2}{\phi_1} X_1.
\end{equation}
Thus, we define $\mathcal{Z}_0^3$ as
\begin{align*}
    \mathcal{Z}_0^3 \triangleq &\{G \in \mathcal{G} \vert G \text{ belongs to Case 3} \} \\ 
    \cap &\{G \in \mathcal{G} \vert G \text{ satisfies \eqref{eq:case_3_max_1}, \eqref{eq:case_3_max_2}, or \eqref{eq:case_3_opp}} \}
\end{align*}
which is a measure-zero subset of $\mathcal{G}$.

\subsubsection*{Case 4}
Finally, consider any game whose pre- and post-transfer parameters strictly satisfy the conditions of Case 4 provided in Table \ref{table1}. It is shown in \cite{kovenock2012coalitional} that there do not exist any opportunities for mutual improvement for any game belonging to Case 4, as the sum of the players' payoffs is maximized in this case. Thus, we define $\mathcal{Z}_0^4 \triangleq \{ G \in \mathcal{G} \vert \text{$G$ belongs to Case 4} \}$. Note that $\mathcal{Z}_0^4$ is also a measure-zero subset of $\mathcal{G}$.

The four cases discussed comprehensively address almost every game $G \in \mathcal{G}$. However, note that the gradients of the payoff functions of the players are not well-defined for games whose parameters lie on the boundary between cases, because the payoff functions are not differentiable along these boundaries. Thus, we do not make any claims regarding the existence of mutually beneficial joint transfers for any games that lie on these boundaries, which are given by
\begin{equation*}
    \mathcal{B}^1 \triangleq \bigg\{G \in \mathcal{G} \bigg\vert \frac{\phi_1}{\phi_2} = \frac{1}{X_1 X_2}\bigg\}, \quad \mathcal{B}^2 \triangleq \bigg\{G \in \mathcal{G} \bigg| 1 - \sqrt{\frac{\phi_1 X_1 X_2}{\phi_2}} = X_2 \bigg\}.
\end{equation*}
However, note that $\mathcal{B}^1$ and $\mathcal{B}^2$ are also measure-zero subsets of $\mathcal{G}$. After considering the boundaries, every game $G \in \mathcal{G}$ has been addressed. Thus, we can write $\mathcal{G}^{b,v}_0$ as
\begin{equation}
    \mathcal{G}^{b,v}_0 = \mathcal{B}^1 \cup \mathcal{B}^2 \cup \mathcal{Z}_0^1 \cup \mathcal{Z}_0^2 \cup \mathcal{Z}_0^3 \cup\ \mathcal{Z}_0^4.
\end{equation}
The subset $\mathcal{G}^{b,v}_0$ is the union of finitely many measure-zero subsets, so it is also a measure-zero subset. \hfill $\square$

\begin{figure*}
    \includegraphics[width=\textwidth]{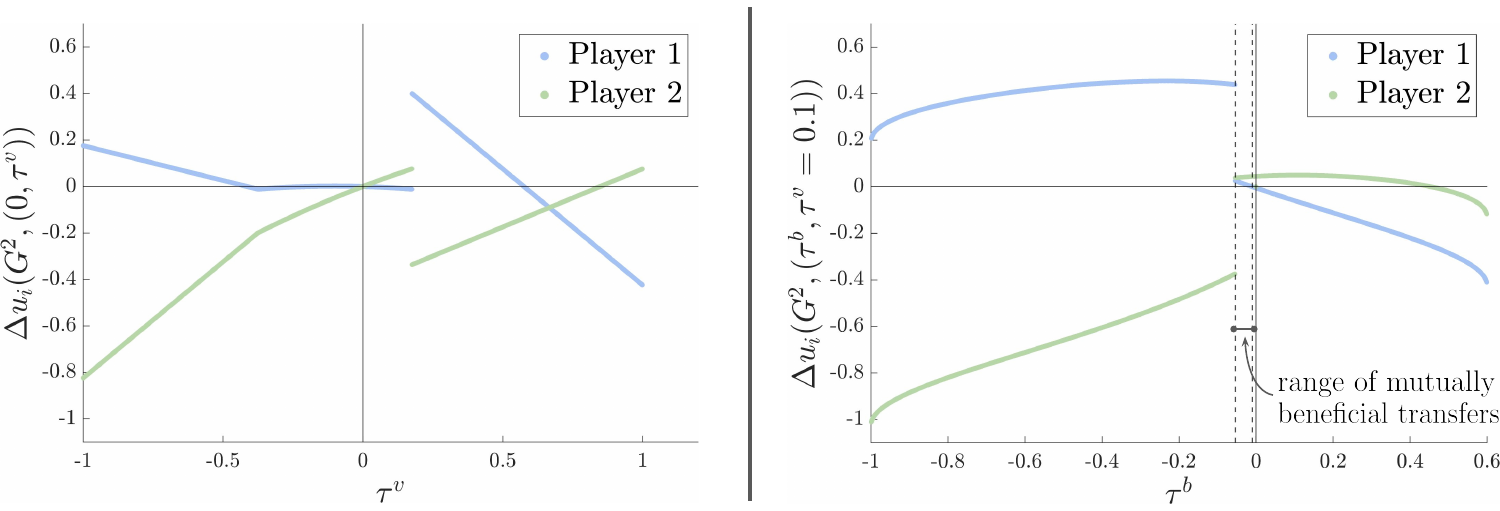}
    \caption{For game $G^2$, the change in payoff of each player as a function of the battlefield transfer (left) and budgetary transfer when $\tau^v = 0.1$ (right). The change in payoff in each plot is shown with respect to the payoff in the original game $G^2$.}
    \label{fig:sim_joint}
\end{figure*}

We call attention to the fact that our proof of the Theorem not only demonstrates the existence of mutually beneficial joint transfers, but also provides a straightforward procedure for identifying them. In particular, players can compute the gradients of their payoff functions, then identify a direction along which both their payoffs increase (e.g., the average of the normalized gradients). By definition, a transfer in this direction is mutually beneficial.

Finally, before proceeding, we make one final observation. In \cite{kovenock2012coalitional}, it is shown that mutually beneficial budgetary transfers always go from the relatively stronger player to the relatively weaker player. However, this is not the case for joint transfers; that is, the constituent battlefield transfer and the constituent budgetary transfer of a mutually beneficial joint transfer need not go from the stronger player to the weaker player, or vice versa.

\subsection{Numerical Example Revisited}

Recall that when players are limited to budgetary transfers, there do not exist any opportunities for mutual improvement in game $G^2$. When players are limited to battlefield transfers instead, the result is similar. The left plot in Figure \ref{fig:sim_joint} shows the change in payoff of each player as a function of the battlefield transfer $\tau^v$. Observe that for any transfer amount, at least one player experiences a lower payoff. We can draw the conclusion that when players consider budgetary \textit{or} battlefield transfers individually in game $G^2$, they cannot do better. However, the main Theorem establishes the existence of mutually beneficial joint transfers for almost all game instances. This naturally leads to the question: If players consider joint transfers instead, do opportunities for mutual improvement arise in game $G^2$?

As one might suspect, such opportunities arise indeed. Suppose that the players perform a battlefield transfer $\tau^v = 0.1$, so that the post-transfer valuations are $\overline{\phi}_1 = 0.9$ and $\overline{\phi_2} = 1.3$ instead. The change in payoff of each player as a function of the budgetary transfer is shown in the right plot in Figure \ref{fig:sim_joint} for game $G^2$ when $\tau^v = 0.1$. Notice that a range of mutually beneficial budgetary transfer opportunities has been reintroduced.

This two-phase interpretation of the joint transfer yields insight into its ubiquity. Figure \ref{fig:example} shows that Player 1 would favor a negative budget transfer, while Player 2 would not; conversely, Figure \ref{fig:sim_joint} shows that Player 2 would favor a positive battlefield transfer, while Player 1 would not. The players can overcome this impasse by performing both transfers simultaneously, thus reaching a configuration that rewards them both. This analysis suggests that in resource allocation games, enhancing the space of strategic maneuvers may present opportunities for compromise that would not be available otherwise.


\section{Conclusion}

In this paper, we considered a three-stage, three-agent coalitional Blotto game where two players compete against an adversary. We allowed the players to transfer both battlefields and budgets in a so-called joint transfer, and we provided sufficient conditions for the existence of mutually beneficial joint transfers, which were shown to exist almost everywhere. In doing so, we demonstrated their merit as a novel method of alliance formation. 

The strategies discussed in this paper present new avenues for exploration in competitive resource allocation settings. Future work includes analyzing the validity and efficacy of these alliance formation strategies using real world examples, such as winner-take-all political elections. Another direction lies in studying these existence of similar alliance formation strategies under different assumptions on the underlying structure of the game; for example, Blotto games are often formulated by parameterizing agents' budgets by per unit production costs instead of predetermined fixed amounts. It would be interesting to analyze whether transferring production costs, battlefields, or a combination of the two would be beneficial in this setting as well.


\begin{acks}
This work is supported by ONR grant \#N00014-20-1-2359 and AFOSR grants \#FA9550-20-1-0054 and \#FA9550-21-1-0203.
\end{acks}


\bibliographystyle{ACM-Reference-Format} 
\bibliography{references}


\end{document}